# Monolayer MXenes: promising half-metals and spin gapless semiconductors

Guoying Gao,[*,a] Guangqian Ding,[a] Jie Li,[a] Kailun Yao,[a] Menghao Wu[a] and Meichun Qian[b]

[a]*School of Physics and Wuhan National High Magnetic Field Center, Huazhong University of Science and Technology, Wuhan 430074, China. E-mail: guoying_gao@mail.hust.edu.cn*

[b]*Department of Physics, Virginia Commonwealth University, Richmond, Virginia 23284, USA*

Half-metals and spin gapless semiconductors are promising candidates for spintronic applications due to the complete (100%) spin polarization of electrons around the Fermi level. Based on recent experimental and theoretical findings of graphene-like monolayer transition metal carbides and nitrides (also known as MXenes), we demonstrate from first-principles calculations that monolayer $Ti_2C$ and $Ti_2N$ exhibit nearly half-metallic ferromagnetism with the magnetic moments of 1.91 and 1.00 $\mu_B$ per formula unit, respectively, while monolayer $V_2C$ is a metal with instable antiferromagnetism, and monolayer $V_2N$ is a nonmagnetic metal. Interestingly, under a biaxial strain, there is a phase transition from nearly half-metal to truly half-metal, spin gapless semiconductor, and metal for monolayer $Ti_2C$. Monolayer $Ti_2N$ is still a nearly half-metal under a proper biaxial strain. Large magnetic moments can be induced by the biaxial tensile and compressive strains for monolayer $V_2C$ and $V_2N$, respectively. We also show that the structures of these four monolayer MXenes are stable according to the calculated formation energy and phonon spectrum. Our investigations suggest that, unlike monolayer graphene, monolayer MXenes $Ti_2C$ and $Ti_2N$ without vacancy, doping or external electric field exhibit intrinsic magnetism, especially the half-metallic ferromagnetism and spin gapless semiconductivity, which will stimulate further studies on possible spintronic applications for new two-dimensional materials of MXenes.

# 1. Introduction

In spintronic applications, high-spin-polarized materials are required for the improvement of the performance of spintronic devices such as spin filter, spin diode, and spin valve.[1,2] Half-metallic magnets (HMMs)[2,3] and spin gapless semiconductors (SGSs)[4] have been expected as the most promising candidates for high-spin-polarized materials due to the complete (100%) spin polarization of electrons around the Fermi level. In the past few years, HMMs and SGSs have been theoretically predicted or experimentally synthesized in some Heusler alloys, metallic oxides, perovskite compounds, transition-metal chalcogenides and pnicides, and so on.[2-9] Interestingly, since the experimental discovery of two-dimensional (2D) graphene,[10] also known as monolayer graphite, some HMMs and SGSs have also been found in graphene and graphene-like monolayers and nanoribbons by selective doping or applying an external electric field.[11-16] In addition, the magnetic properties and half-metallicity were predicted in the 2D transition-metal phthalocyanine monolayers and the metal-organic frameworks by changing the metal centers with different transition-metals.[17,18] So, It is meaningful to search for HMMs and SGSs in some new 2D materials due to tunable physical properties.

Recently, a new class of graphene-like 2D materials of transition-metal carbides/nitrides, also known as MXenes, have been realized experimentally.[19,20] MXenes can be achieved by selective etching of the A atomic layer from the MAX phases with the chemical composition of $M_{n+1}AX_n$ (n=1, 2, or 3). Here, M is an early transition metal such as Ti and V, A is an A-group (mostly IIIA and IVA) element such as Al and Si, and X is C or N. MAX phases are layered hexagonal compounds, and more than 70 MAX phases have been realized experimentally.[21,22] In the MAX phase, the M-X-M bond is the mixture of metallic, covalent and ionic, and the bond strength is strong. But the M-A bond is metallic and weaker than the M-X-M bond. This bond characteristic and the layered structure in MAX phases make it feasible to remove the A atomic layer from MAX phases, forming the 2D MXenes. The first discovered MXene is $Ti_3C_2$, which was synthesized in 2011 by immersing $Ti_3AlC_2$ powers in hydrofluoric acid at room temperature.[19] So far, more and more MXenes were predicted theoretically, and several of them have been realized experimentally, e.g., $Ti_3C_2$, $Ti_2C$, $Nb_2C$, $V_2C$, $Ta_4C_3$, $Nb_4C_3$, $TiNbC$, $(V_{0.5}Cr_{0.5})_3C_2$, and $Mo_2C$.[19,20,23-28] MXenes have attracted extensive interest, because they were found to be promising candidates for the applications in electrode materials, supercapacitors, hybrid devices, optical devices, thermoelectric materials, and even topological insulators.[29-37]

However, there are few reports on the magnetic properties of MXenes up to date.[35,38-40] Khazaei *et al.* proposed that F- and OH-surface-functionalized MXenes $Cr_2C$ and $Cr_2N$ are ferromagnetic.[35] Si *et al.* predicted from first-principles that monolayer MXene $Cr_2C$ exhibits HM ferromagnetism.[39] However, both $Cr_2C$ and $Cr_2N$ MXenes have not been realized in experiment till now. Recently, Zhao *et al.*[40] studied the strain effect on the magnetic properties of monolayer MXenes $M_2C$ (M=Hf, Nb, Sc, Ta, Ti, V, Zr). But they only calculated the change of magnetic moment under strain for $M_2C$, and only the electronic structure of $Hf_2C$ was discussed. No HMMs or SGSs were found due to the lack of the analysis of electronic structure for all MXenes studied. In this paper, we systemically investigate the electronic structure, magnetic properties, and structural stability for four monolayer MXenes $M_2X$ (M=Ti,V and X=C,N), where $Ti_2C$ and $V_2C$ have been synthesized experimentally.[20,23] We demonstrate from first-principles calculations that nearly half-metallicity can be obtained in monolayer $Ti_2C$ and $Ti_2N$, and there is an interesting phase transition from nearly half-metal to truly half-metal, SGS, and metal for monolayer $Ti_2C$ under a biaxial strain, which means monolayer Ti$_2$C and Ti$_2$N without vacancy or doping would be useful in spintronic applications.

## 2. Results and discussion

**2.1 Magnetism and electronic structure at equilibrium lattice**

Fig. 1 presents the side and top views of the structure of monolayer MXene M$_2$X. To achieve the monolayer MXene, we firstly remove the A atoms from the bulk MAX phase, then cleave the (001) surface, and add a 20 Å vacuum to the two adjacent atomic layers to avoid the interactions each other. Similar to monolayer MoS$_2$, monolayer MXene M$_2$X has a hexagonal symmetry, and one X atomic layer is sandwiched by two M atomic layers. In this study, the 2×2 supercell of the unit cell of M$_2$X monolayer is adopted.

Before calculating the electronic structure and magnetic properties of monolayer M$_2$X, we perform the spin-polarized geometry optimization by relaxing both lattices and atomic positions. We show the optimized lattice constants of monolayer M$_2$X in Table 1. Our results are in good agreement with previous works except for those from ref. 42 and ref. 43 (see Table 1). The difference is that the spin-polarization was not considered in ref. 42 and ref. 43. Our optimized lattice constant without spin polarization is 3.035 Å for Ti$_2$C, which is indeed close to those from ref. 42 and ref. 43. So, the comparable lattice constants shown in Table 1 indicate the reliability of our present calculations. To determine the possible magnetism and the magnetic ground state for monolayer M$_2$X, we calculate the total energies with ferromagnetic (FM), antiferromagnetic (AFM), and nonmagnetic (NM) states, which are listed in Table 1. For the AFM state, similar to monolayer Cr$_2$C,[39] two possible configurations are considered, which are shown in Fig. 1(c) and (d). One can see from Table 1 that the FM states have lowest total energies than the AFM and NM ones for both $Ti_2C$ and $Ti_2N$, and thus the FM states are most stable, especially the total energy differences between FM and AFM1, and between FM and AFM2, respectively, reach -35 meV and -74 meV per formula unit for $Ti_2C$. The large energy difference means that $Ti_2C$ has a high Curie temperature. We also note that the AFM1 states have lower total energies than the AFM2 states for both $Ti_2C$ and $Ti_2N$, the reason is that the Ti-Ti atomic distance in AFM1 state is shorter than that in AFM2 state, leading to the stronger AFM interaction in AFM1 state than in AFM2 state. For $V_2C$, the AFM1 state has lowest total energy than the FM and AFM2 ones, i.e., $V_2C$ is AFM, this is in agreement with the previous work.[44] However, we note that the total energy differences among the four magnetic states for $V_2C$ are very small, which means the AFM state of $V_2C$ is instable under certain circumstance. For $V_2N$, the total energies are the same for all four possible magnetic states, and thus $V_2N$ is NM.

Table 1 shows the calculated total and atomic magnetic moments for four monolayer $M_2X$ with FM state. It is clearly that both $Ti_2C$ and $Ti_2N$ have considerable magnetic moments of 1.91 and 1.00 $\mu_B$ per formula unit, respectively, which mainly come from Ti atom. The magnetic moments of C and N atoms have very small negative values, which originate from the interactions of C and N atoms with neighboring Ti atoms. The calculated total magnetic moment of 1.91 $\mu_B$ per formula unit for $Ti_2C$ is in good agreement with those of 1.92 $\mu_B$ and 1.85 $\mu_B$ obtained by Zhao et al.[40] and Xie et al.[38] respectively. For $V_2C$, the total magnetic moment of 0.14 $\mu_B$ per formula unit is very small, which is consistent with the previous work.[40] The calculated total and atomic magnetic moments are all zero, indicating the nonmagnetism of $V_2N$, which is in agreement with aforementioned total energy calculations: $V_2N$ has the same total energy within

FM, AFM and NM states.

In order to search for possible HMMs and SGSs in four monolayer M$_2$X, we calculate the spin-resolved band structures at their equilibrium lattices with FM state, which are shown in Fig. 2. One can see clearly that both Ti$_2$C and Ti$_2$N exhibit nearly half-metallicity, i.e., one of the two spin channels is metallic, and the other is nearly semiconducting, because the bottom of the conduction band (minority-spin conduction band for Ti$_2$C and majority-spin conduction band for Ti$_2$N) touches the Fermi level a little. Both the minority-spin energy gap for Ti$_2$C and the majority-spin energy gap for Ti$_2$N are very small. It should be pointed out that the nearly half-metallicities of monolayer Ti$_2$C and Ti$_2$N are intrinsic, which is superior to those of graphene and graphene-like monolayers or nanoribbons, where the emergence of half-metallicity requires a selective doping or an external electric field.[11-16] The calculated total magnetic moments (see Table 1), nearing to the integer of Bohr magneton,[2,7] also reflect the characteristic of HM ferromagnetism in Ti$_2$C and Ti$_2$N. For V$_2$C, as shown in Fig. 2(c), it is metallic because the electrons in both spin channels cross the Fermi level. But the spin polarization of the band structure is very week, which indicates that the magnetic moment in V$_2$C should be small (see Table 1). For the case of V$_2$N, it is also a metal, but there is no spin polarization of the band structure, and thus it is a NM metal.

In order to understand in details the electronic structure and magnetism in monolayer M$_2$X, we show in Fig. 3 the calculated total and main partial density of states (DOS) of monolayer Ti$_2$N at equilibrium lattice as an example of monolayer M$_2$X. For both spin channels, the total DOS can be divided into three parts: the first one locates at the lowest energy region around -5 eV below the Fermi level, the second one is between -2 eV and the Fermi level, and the third one is above the Fermi level. It is clear that the first part of the total DOS mainly consists of N 2$p$ states with a small mixture of Ti 3$d$ states, which form the six fully filled bands in both spin channels by six electrons. Both the second and third parts of total DOS are mainly from Ti 3$d$ states with a small contribution from N 2$p$ states. The second part forms three majority-spin fully filled bands, one minority-spin fully filled band, and two partially filled minority-spin bands (see Fig. 2(b)), which are occupied by five electrons mainly from Ti 3$d$ electrons with a little N 2$p$ electrons, including three majority-spin electrons and two minority-spin ones, which results in the total magnetic moment of 1.00 $\mu_B$ per formula unit in Ti$_2$N, and the total magnetic moment is mainly carried by the Ti atom. As seen from Fig. 3, similar to monolayer Cr$_2$C,[36] the large exchange splitting of Ti 3$d$ electrons and the strong hybridization of Ti 3$d$ electrons with N 2$p$ electrons are responsible for the formation of the HM ferromagnetism with majority-spin energy gap in monolayer Ti$_2$N.

**2.2 Strain effect**

Let us now discuss the strain effect on the electronic structure and magnetism of monolayer M$_2$X. This is feasible, because both experimental and theoretical studies have indicated that the electronic and magnetic properties are tunable by the strain for graphene and graphene-like 2D materials such as monolayer transition-metal dichalcogenides.[45-47] Here, we apply a biaxial strain to the monolayer M$_2$X, which is defined as $\varepsilon = (a - a_0)/a_0$, where a and a$_0$ are the lattice constants of the strained and the equilibrium unit cells, respectively. Positive and negative a$_0$ correspond to the tensile and compressive strains, respectively. Very recent studies by Guo *et al.*[48] have indicated that the structure of monolayer Ti$_2$C with a large strain (up to 9.5% biaxial tensile strain) is stable, but they did not investigate the magnetic properties and the spin-polarized electronic structure. Here, we consider the biaxial strain from -5% to 5% for monolayer M$_2$X, and the atomic positions are fully relaxed. Fig. 4 gives the calculated total magnetic moments as a function of strain. One can see that the change of the magnetic moment of monolayer $Ti_2C$ is very small, which means the spin splitting of Ti 3d electrons has no obvious change under the biaxial strain. As can be seen in Fig. 5, both the majority-spin and minority-spin bands shift towards higher energy regions under strain for $Ti_2C$. The change of magnetic moment of $Ti_2N$ is similar to that of $Ti_2C$, however, under a larger tensile strain (above 3%), the magnetic moment of $Ti_2N$ increases clearly, the reason is that the spin splitting of the bands around the Fermi level from Ti $3d$ states becomes strong with the increase of strain. Remarkably, for $V_2C$ and $V_2N$ with small and zero magnetic moments at strain-free state, respectively, magnetic moments are induced by larger tensile or compressive strains. In particular, the magnetic moment of $V_2C$ is increased to about 1.0 $\mu_B$ per formula unit by applying a biaxial tensile strain of 10% (not shown in Fig. 4), which is in agreement with the result by Zhao *et al.*[40] The magnetic moment reaches a maximum of 1.2 $\mu_B$ at the strain state of 12%, and then decreases with further increase of the tensile strain for $V_2C$. Similarly, the magnetic moment of $V_2N$ reaches a maximum of 0.9 $\mu_B$ at the strain state of -7%, and then decreases with further increase of the compressive strain. All the changes of magnetic moment of four monolayer $M_2X$ under strain are dominated by the Ti and V atoms, while the X atomic magnetic moments have very small changes under strain.

We now analysis the change of electronic band structure under strain. By comparing the band structures under different strains, we find that there is an interesting transition from a half-metal to a SGS and a metal for monolayer $Ti_2C$. The phase diagram of electronic properties and band structure are summarized in Figs. 4-6. The nearly HM characteristic of monolayer $Ti_2C$ can be retained under the biaxial strains from -3% to 2%, and especially the truly half-metallicity with a minority-spin gap of 0.04 eV emerges at the strain state of -2% (see Fig. 5(b)). HM monolayer Ti$_2$C becomes a metal under a compressive strain larger than -3%, and becomes a SGS under a tensile strain larger than 2%. Note that this is the first finding of SGS in MXenes. As shown in Fig. 5(d), the band structure is spin polarized, but the energy gap is zero. Compared to HMMs, SGSs have also the complete spin polarization of electrons around the Fermi level on one hand, and on another hand, no energy is required to excite electrons from the valence band to the conduction band in SGSs, which means the spintronic devices based on SGSs would possess small energy consumption.[4,49] For the case of monolayer Ti$_2$N, it is still a nearly half-metal under the biaxial strains from -3% to 3%, but it becomes a metal with both larger tensile and compressive strains (see Fig. 5). For both V$_2$C and V$_2$N, as shown in Fig. 6, although the band structures are spin polarized under larger biaxial tensile and compressive strains, respectively, they are still metals under the biaxial strains from -5% to 5%. Therefore, among the four monolayer MXenes, the properties of half-metals and SGSs of monolayer Ti$_2$C and Ti$_2$N would be useful in spintronic applications.

**2.3 Structural stability**

The previous experimental works have indicated that the surfaces of monolayer MXenes are usually not pristine, but terminated with functional groups such as F, OH, O or even the mixture of them.[25] The different surface functional groups can tune the physical properties of monolayer MXenes,[35-39] and interestingly, some pristine MXenes have been predicted to be promising anode materials with high storage capacity than those with surface functional groups.[29,34] So, it is meaningful to synthesize pristine MXenes experimentally as expected by Naguib et al.[25] Xie et *al.*[34] have proposed a feasible strategy for the synthesis of pristine MXenes. Remarkably, a recent experiment has confirmed that the surface functional groups F and OH in monolayer $Ti_2C$ can be eliminated by heat treatment at different temperatures.[50] This is just the reasonability that we study here pristine monolayer $M_2X$ instead of those with surface functional groups. As far as we know, however, there have been no experimental reports on the elimination of surface functional groups for monolayer MXenes except for $Ti_2C$.[50] Therefore, it is necessary to assess and compare the structural stability of pristine monolayer $M_2X$ studied here. On one hand, we calculate the cohesive energy by the formula expressed by: $E_c = E_{M_2X} - 2E_M - E_X$, where $E_{M_2X}$, $E_M$, and $E_X$ are the total energies of monolayer $M_2X$, and free atoms of M and X, respectively. Our calculated cohesive energies for the four monolayer MXenes of $Ti_2C$, $Ti_2N$, $V_2C$, and $V_2N$ are -19.92, -17.20, -18.78, and -15.63 eV per formula unit, respectively. These large negative cohesive energies indicate the feasibility of the formation of monolayer $Ti_2C$, $Ti_2N$, $V_2C$, and $V_2N$. On another hand, we calculate the formation energies. (i) There have been experimental MAX phases of $Ti_2AlC$, $Ti_2GaN$, $V_2AlC$ and $V_2GaN$,[21,22] and thus we consider the balance equations: $Ti_2AlC - Al \rightarrow Ti_2C$, $Ti_2GaN - Ga \rightarrow Ti_2N$, $V_2AlC - Al \rightarrow V_2C$, and $V_2GaN - Ga \rightarrow V_2N$. The formation energy of monolayer $M_2X$ can be expressed by $E_{f1} = E_{M_2X} - E_{M_2AX} + E_X$, where $E_{M_2X}$, $E_{M_2AX}$, and $E_X$ are the total energies of monolayer $M_2X$, MAX phases, and C or N, respectively. The calculated formation energies are 2.69, 3.15, 2.93, and 2.83 eV per formula unit for $Ti_2C$, $Ti_2N$, $V_2C$, and $V_2N$, respectively.[51] These values are comparable, and thus it would be feasible to obtain monolayer $Ti_2N$, $V_2C$, and $V_2N$ by etching A atomic layer from the corresponding MAX phases like experimental monolayer $Ti_2C$ from $Ti_2AlC$. We note that our present results of 2.69 eV for $Ti_2C$ and 3.15 eV for $Ti_2N$ are in good agreement with previous theoretical values of 2.78 eV and 3.00 eV, respectively.[42] (ii) In order to check if monolayer $M_2X$ will be decomposed into M and X or not, we calculate the formation energy according to the equation $E_{f2} = E_{M_2X} - 2E_M - E_X$, where $E_{M_2X}$ is the total energy of monolayer $M_2X$, and $E_M$ and $E_X$ are the total energies of M and X, respectively. The calculated formation energies are -0.25, -2.20, -0.88, and -0.65 eV per formula unit for monolayer $Ti_2C$, $Ti_2N$, $V_2C$, and $V_2N$, respectively. The negative formation energies indicate that monolayer $M_2X$ should be stable and would not be decomposed into M and X.

Finally, we calculate the phonon spectra of four monolayer $M_2X$ at equilibrium lattices, which are shown in Fig. 7. Clearly, these phonon spectra are similar, and importantly, all the spectra have no imaginary frequencies, which mean that the structures of the four monolayer MXenes are stable. Therefore, in addition to the experiment pristine monolayer $Ti_2C$,[50] our present prediction of monolayer $Ti_2N$, $V_2C$, and $V_2N$ without surface functional groups would be realized by future experiments. It is also interesting to study the magnetic properties such as half-metallicity and spin gapless semiconductivity in future experiments. Certainly, as mentioned above, it is difficult to eliminate the surface functional groups to some extent, and it needs some experimental techniques such as heat treatment at different temperatures or other synthesis methods.

## 3. Conclusion

In summary, recently discovered graphene-like 2D materials of monolayer transition metal carbides and nitrides (also known as MXenes), which can be obtained by etching A atomic layer from the MAX phases, have attracted much attention due to excellent electrical and electrochemical properties, but the studies on their magnetic properties are few. In this paper, we have used the first-principles calculations to investigate systemically the structural, electronic, and magnetic properties of four monolayer $M_2X$. We find that pristine monolayer $Ti_2C$ and $Ti_2N$ behave nearly half-metallicity, and there is an interesting phase transition from nearly half-metal to truly half-metal, spin gapless semiconductor, and metal for monolayer $Ti_2C$ under a biaxial strain. Large magnetic moments can be induced by biaxial tensile and compressive strains for monolayer $V_2C$ and $V_2N$, respectively. We also predict that these four monolayer $M_2X$ are stable from the calculated formation energy and the phonon spectrum. The present study suggests that the magnetism of pristine monolayer $M_2X$ can be induced/tuned by strain, and the half-metallicity and the spin gapless semiconductivity of $Ti_2C$ and $Ti_2N$ would be useful in spintronic applications.

## 4. Computational methods

The structural optimization, electronic structure, and magnetic properties of monolayer MXenes $M_2X$ (M=Ti,V and X=C,N) are calculated by the first-principles full-potential linearized augmented plane-wave method (Wien2k package).[52] In this method, the generalized gradient approximation (GGA) in the Perdew-Burke-Ernzerhof (PBE) scheme[53] for the electronic exchange-correlation functional is adopted, and relativistic effect in the scalar approximation is considered. The radii $R_{mt}$ of the muffin tins are chosen to be approximately proportional to the corresponding ionic radii and as large as possible under the condition that the spheres do not overlap. We take $R_{mt}K_{max}$ relating to the energy cutoff equal to 8.5 and make the expansion up to $l$=10 in the muffin tins. The $20 \times 20 \times 1$ $k$ meshes in the Brillouin zone integration are used. The self-consistency calculations are considered to be achieved when the total energy difference between succeeding iterations is less than $10^{-5}$ Ry per formula unit.

It should be pointed out that we also use the hybrid (HSE06) functional[54] implemented in the VASP code[55] to calculate the spin-polarized band structure of monolayer MXenes at equilibrium lattices, and find that the band structures (see Fig. 8) are similar to those based on the GGA-PBE (see Fig. 2), and both monolayer $Ti_2C$ and $Ti_2N$ are still HM. Remarkably, the majority-spin gap within hybrid (HSE06) functional increases significantly compared to that obtained by GGA-PBE for monolayer $Ti_2N$, this is due to the fact that the hybrid functional usually correct the underestimation of the energy gap from GGA-PBE. In addition, we also find that it is still feasible to induce SGS for monolayer $Ti_2C$ by appropriate biaxial strain when the hybrid functional is considered.

For the calculations of phonon spectrum of monolayer $M_2X$, we employ the finite displacement method implemented in the CASTEP code[56] based on the first-principles pseudopotential plane-wave method within GGA-PBE.[53] The ultrasoft pseudopotentials for M and X atoms are adopted for the interactions of electrons with ion cores, and the energy cutoff used is 500 eV. Same to the electronic structure calculations, the $20 \times 20 \times 1$ $k$ meshes in the Brillouin zone integration are also adopted in the calculations of phonon spectrum. The CASTEP code has been successfully used to calculate the phonon spectrum of some 2D materials.[57,58]

**Acknowledgment**

We thank Y. Du and Z. Zhou for useful discussions. This work was supported by the National Natural Science Foundation of China under Grant No. 11474113, by the Natural Science Foundation of Hubei Province under Grant No. 2015CFB419, and by the Fundamental Research Funds for the Central Universities under Grant No. HUST: 2015TS019.

**Notes and references**

1. I. Zutic, J. Fabian and S. S. Sarma, *Rev. Mod. Phys.*, 2004, **76**, 323.

2. C. Felser, G. H. Fecher and B. Balke, *Angew. Chem., Int. Ed.*, 2007, **46**, 668-699.

3. R. A. de Groot, F. M. Mueller, P. G. van Engen and K. H. J. Buschow, *Phys. Rev. Lett.*, 1983, **50**, 2024.

4. X. L. Wang, *Phys. Rev. Lett.*, 2008, **100**, 156404.

5. T. Graf, C. Felser and S. P. P. Parkin, *Prog. Solid State Chem.*, 2011, **39**, 1-50.

6. S. Ouardi, G. H. Fecher, C. Felser and J. Kubler, *Phys. Rev. Lett.*, 2013, **110**, 100401.

7. G. Y. Gao and K.-L. Yao, *Appl. Phys. Lett.*, 2013, **103**, 232409.

8. M. C. Qian, C. Y. Fong, W. E. Pickett, J. E. Pask, L. H. Yang and S. Dag, *Phys. Rev. B: Condens. Matter* 2005, **71**, 012414.

9. S. Skaftouros, K. Ozdogan, E. Sasioglu and I. Galanakis, *Appl. Phys. Lett.*, 2013, **102**, 022402.

10. K. S. Novoselov, A. K. Geim, S. V. Morozov, D. Jiang, Y. Zhang, S. V. Dubonos, I. V. Gregorieva, A. A. Firsov, *Science*, 2004, **306**, 666-669.

11. Y. Li, Z. Zhou, P. Shen and Z. Chen, *ACS Nano*, 2009, **3**, 1952-1958.

12. J. Hu, Z. Zhu and R. Wu, *Nano Lett.*, 2015, **15**, 2074-2078.

13. W. H. Wu, X. J. Wu, Y. Gao and X. C. Zeng, *Appl. Phys. Lett.*, 2009, **94**, 223111.

14. Y. Zhou, Z. Wang, P. Yang, Z. Zu, L. Yang, X. Sun and F. Gao, *ACS Nano*, 2012, **6**, 9727-9736.

15. L. Kou, C. Tang, Y. Yang, T. Heine and C. Chen, *J. Phys. Chem. Lett.*, 2012, **3**, 2934-2941.

16. X. Hu, W. Zhang, L. Sun and A. V. Krasheninnikov, *Phys. Rev. B: Condens. Matter*, 2012, **86**,

195418.

17. J. Zhou and Q. Sun, *J. Am. Chem. Soc.*, 2011, **133**, 15113-15119.

18. Q. Zhang, B. Li and L. Chen, *Inorg. Chem.*, 2013, **52**, 9356-9362.

19. M. Naguib, M. Kurtoglu, V. Presser, J. Lu, J. Niu, M. Heon, L. Hultman, Y. Gogotsi and M. W. Barsoum, *Adv. Mater.*, 2011, **23**, 4248-4253.

20. M. Naguib, O. Mashtalir, J. Carle, V. Presser, J. Lu, L. Hultman, Y. Gogotsi and M. W. Barsoum, *ACS Nano*, 2012, **6**, 1322-1331.

21. P. Eklund, M. Beckers, U. Jansson, H. Hogberg and L. Hultman, *Thin Solid Films*, 2010, **518**, 1851-1878.

22. Z. M. Sun, *Int. Mater. Rev.*, 2011, **56**, 143-166.

23. M. Naguib, J. Halim, J. Lu, K. M. Cook, L. Hultman, Y. Gogotsi and M. W. Barsoum, *J. Am. Chem. Soc.*, 2013, **135**, 15966-15969.

24. M. Ghidiu, M. Naguib, C. Shi, O. Mashtalir, L. M. Pan, B. Zhang, J. Yang, Y. Gogotsi, S. J. L. Billinge and M. W. Barsoum, *Chem. Commun.*, 2014, **50**, 9517-9520.

25. M. Naguib, V. N. Mochalin, M. W. Barsoum and Y. Gogotsi, *Adv. Mater.*, 2014, **26**, 992-1005.

26. K. J. Harris, M. Bugnet, M. Naguib, M. W. Barsoum and G. R. Goward, *J. Phys. Chem. C*, 2015, **119**, 13713-13720.

27. R. Meshkian, L.-A. Naslund, J. Halim, J. Lu, M. W. Barsoum and J. Rosen, *Scripta Mater.*, 2015, **108**, 147-150.

28. Q. Tang and Z. Zhou, *Prog. Mater. Sci.*, 2013, **58**, 1244-1316.

29. Q. Tang, Z. Zhou and P. Shen, *J. Am. Chem. Soc.*, 2012, **134**, 16909-16916.

30. X. Zhang, X. Zhao, D. Wu, Y. Jing and Z. Zhou, *Nanoscale*, 2015, **7**, 16020-16025.

31. Y. Dall'Agnese, P.-L. Taberna, Y. Gogotsi and P. Simon, *J. Phys. Chem. Lett.*, 2015, **6**, 2305-2309.

32. C. Eames and M. S. Islam, *J. Am. Chem. Soc.*, 2014, **136**, 16270-16276.

33. Y. Xie, M. Naguib, V. N. Mochalin, M. W. Barsoum, Y. Gogotsi, X. Yu, K.-W. Nam, X.-Q. Yang, A. I. Kolesnikov and P. R. C. Kent, *J. Am. Chem. Soc.*, 2014, **136**, 6385-6394.

34. Y. Xie, Y. Dall'Agnese, M. Naguib, Y. Gogotsi, M. W. Barsoum, H. L. Zhuang and P. R. C. Kent, *ACS Nano.*, 2014, **8**, 9606-9615.

35. M. Khazaei, M. Arai, T. Sasaki and C.-Y. Chung, *Adv. Funct. Mater.*, 2013, **23**, 2185-2192.

36. M. Khazaei, M. Arai, T. Sasaki, M. Estili and Y. Sakka, *Phys. Chem. Chem. Phys.*, 2014, **16**, 7841-7849.

37. H. Weng, A. Ranjbar, Y. Liang, Z. Song, M. Khazaei, S. Yunoki, M. Arai, Y. Kawazoe, Z. Fang and X. Dai, *Phys. Rev. B: Condens. Matter*, 2015, **92**, 075436.

38. Y. Xie and P. R. C. Kent, *Phys. Rev. B: Condens. Matter*, 2013, **87**, 235441.

39. C. Si, J. Zhou and Z. Sun, *ACS Appl. Mater. Interfaces*, 2015, **7**, 17510-17515.

40. S. Zhao, W. Kang and J. Xue, *Appl. Phys. Lett.*, 2014, **104**, 133106.

41. L.-Y. Gan, Y.-J. Zhao, D. Huang and U. Schwingenschlogl, *Phys. Rev. B: Condens. Matter*, 2013, **87**, 245307.

42. I. R. Shein and A. L. Ivanovskii, *Comput. Mater. Sci.*, 2012, **65**, 104-114.

43. M. Kurtoglu, M. Naguib, Y. Gogotsi and M. W. Barsoum, *MRS Commun.*, 2012, **2**, 133-137.

44. J. Hu, B. Xu, C. Ouyang, S. A. Yang and Y. Yao, *J. Phys. Chem. C*, 2014, **118**, 24274-24281.

45. N. Ferralls, R. Maboudian and C. Carraro, *Phys. Rev. Lett.*, 2008, **101**, 15681.

46. L. Kou, A. Du, C. Chen and T. Frauenheim, *Nanoscale*, 2014, **6**, 5156-5161.

47. H. J. Conley, B. Wang, J. I. Ziegler, R. F. J. Haglund, S. T. Pantelides and K. I. Bolotin, *Nano Lett*., 2013, **13**, 3626-3630.

48. Z. Guo, J. Zhou, C. Si and Z. Sun, *Phys. Chem. Chem. Phys*., 2015, **17**, 15348-15354.

49. X.-L. Wang, S. X. Dou and C. Zhang, *NPG Asia Mater*., 2010, **2**, 31-38.

50. J. Li, Y. Du, C. Huo, S. Wang and C. Cui, *Ceram. Int*., 2015, **41**, 2631-2635.

51. It should be pointed out that we can select different MAX phases for the calculations of formation energies of $M_2X$, but the differences are very small, e.g., monolayer $V_2N$ can be obtained from $V_2GaN$ or $V_2AlN$, and the calculated formation energies are 2.83 eV and 2.86 eV, respectively.

52. P. Blaha, K. Schwarz, G. K. H. Madsen, D. Kvasnicka and J. Luitz, *Comput. Phys. Commun*., 1990, **59**, 399-415.

53. J. P. Perdew, K. Burke and M. Ernzerhof, *Phys. Rev. Lett*., 1996, **77**, 3865.

54. J. Heyd, G. E. Scuseria and M. Ernzerhof, *J. Chem. Phys*. 2006, **124**, 219906.

55. G. Kresse and J. Furthmüller, *Phys. Rev. B: Condens. Matter*, 1996, **54**, 11169.

56. S. J. Clark, M. D. Segall, C. J. Pickard, P. J. Hasnip, M. I. J. Probert, K. Refson and M. C. Payne, *Z. Kristallogr*., 2005, **220**, 567-570.

57. L. Kou, Y. Ma, X. Tian, T. Frauenheim, A. Du and S. Smith, *J. Phys. Chem. C*, 2015, **119**, 6918-6922.

58. T. Hu, M. Hu, Z. Li, H. Zhang, C. Zhang, J. Wang and X. Wang, *J. Phys. Chem. A*, 2015, **119**, 12977-12984.

**Tables**

**Table 1** The optimized lattice constants $a_0$ (in Å) with FM state, the total ($M_{tot}$), atomic and interstitial regional magnetic moments (in $\mu_B$) at equilibrium lattices, and the total energy differences of FM with AFM ($E_{FM-AFM}$) and FM with NM ($E_{FM-NM}$) states (in meV/f.u.) for monolayer $M_2X$. For comparison, available values from other works are also presented.

|  | $a_0$ | $M_{tot}$ | $M_{Ti/V}$ | $M_{C/N}$ | $M_{inter}$ | $E_{FM-AFM1}$ | $E_{FM-AFM2}$ | $E_{FM-NM}$ |
|---|---|---|---|---|---|---|---|---|
| $Ti_2C$ | 3.083 | 1.91 | 0.96 | -0.04 | 0.03 | -35 | -74 | -118 |
|  | 3.078 (ref. 40) | | | | | | | |
|  | 3.076 (ref. 41) | | | | | | | |
|  | 3.0395 (ref. 42) | | | | | | | |
|  | 3.007 (ref. 43) | | | | | | | |
| $Ti_2N$ | 2.981 | 1.00 | 0.50 | -0.02 | 0.02 | -7 | -34 | -51 |
|  | 2.9853 (ref. 42) | | | | | | | |
| $V_2C$ | 2.901 | 0.16 | 0.08 | -0.01 | 0.01 | 0.16 | 0.07 | -0.4 |
|  | 2.897 (ref. 40) | | | | | | | |
|  | 2.869 (ref. 43) | | | | | | | |
| $V_2N$ | 2.897 | 0.00 | 0.00 | 0.00 | 0.00 | 0 | 0 | 0 |

**Figures with Figures Captions**

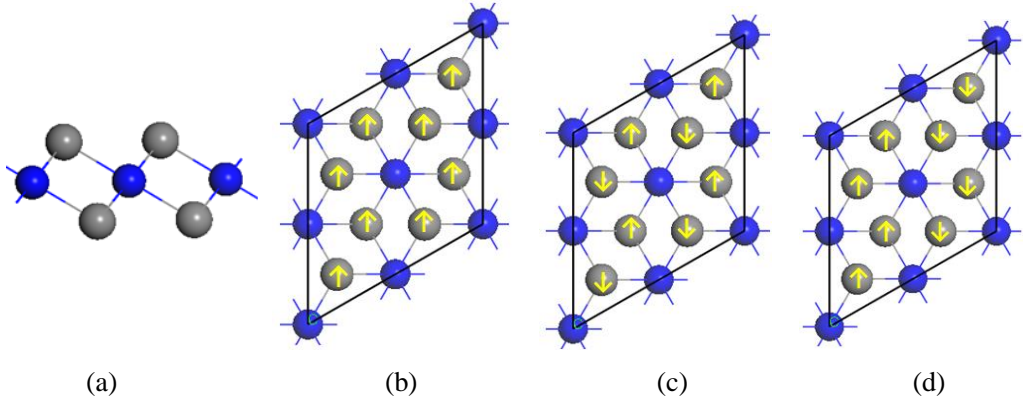

      (a)                    (b)                    (c)                    (d)

**Fig. 1** (a) Slid view of monolayer $M_2X$, and (b-d) top views of monolayer $M_2X$ with ferromagnetic (FM) and antiferromagnetic (AFM1 and AFM2) states. Grey and blue balls represent the M and X atoms, respectively. Yellow arrows represent the spin directions of M atoms. The 2×2 supercell is used.

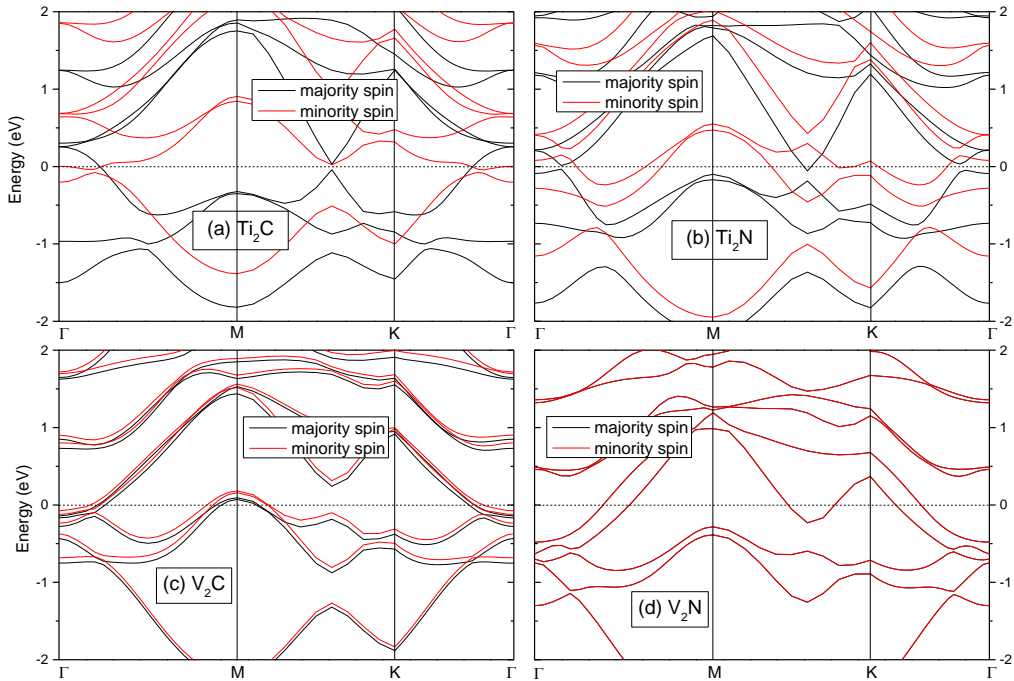

**Fig. 2** Spin-resolved band structure of monolayer $M_2X$ (M=Ti, V and X=C, N) at the equilibrium lattice constants. The dashed line indicates the Fermi level at zero eV.

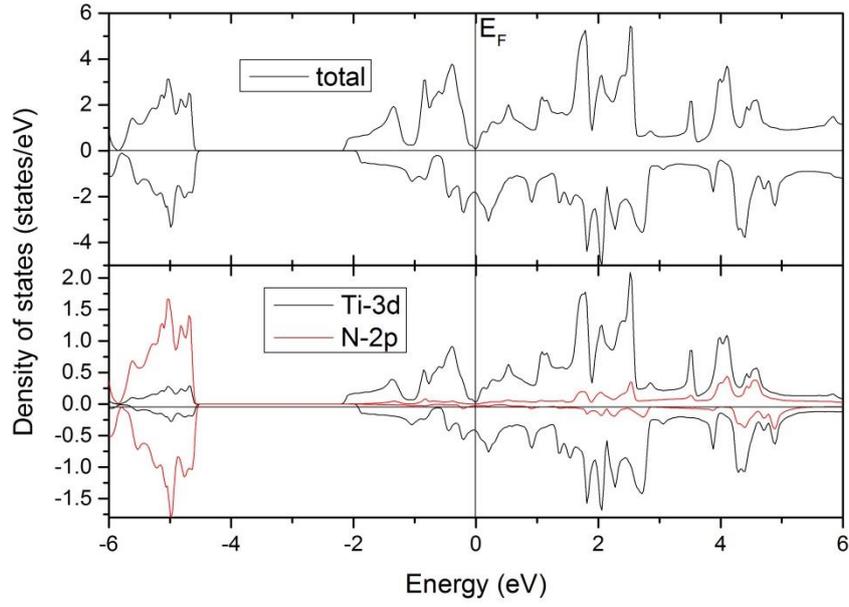

**Fig. 3** Spin-resolved total and main partial density of states of monolayer Ti$_2$N at equilibrium lattice constant.

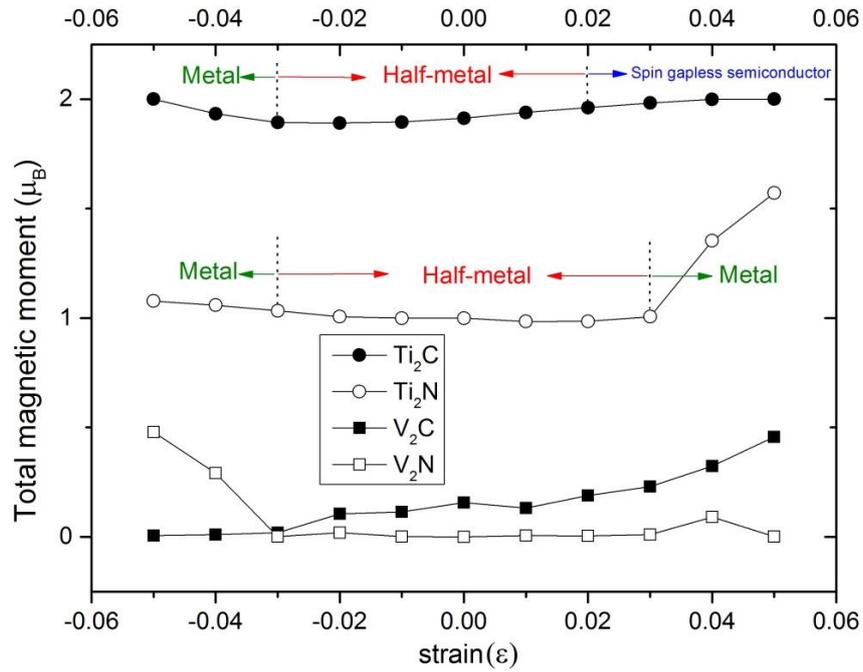

**Fig. 4** Calculated total magnetic moments (in μ$_B$ per formula unit) of monolayer M$_2$X under biaxial strain. For Ti$_2$C and Ti$_2$N, we also give the different physical properties under different strains. Both V$_2$C and V$_2$N are metallic under all the strains studied.

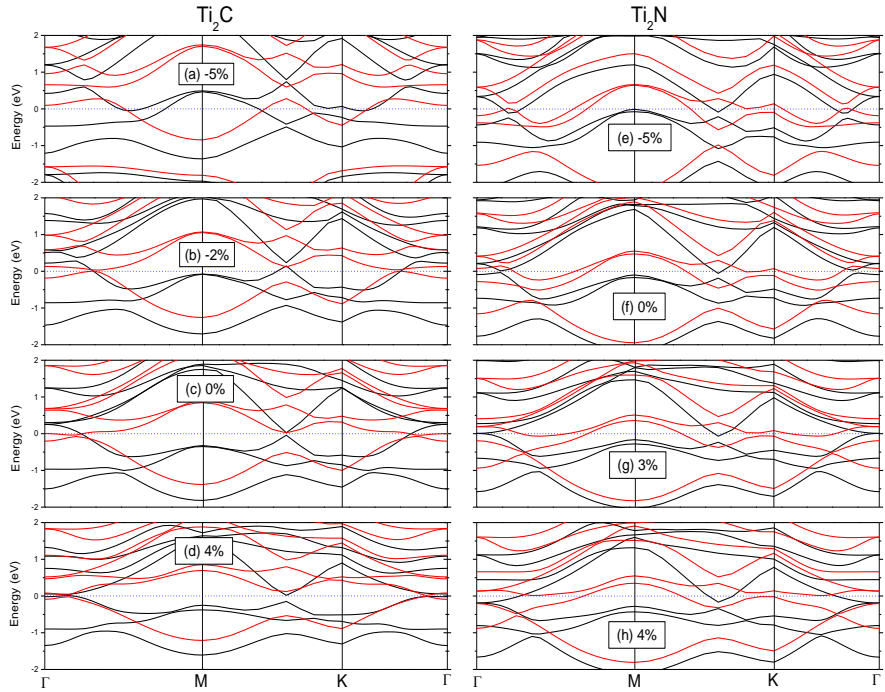

**Fig. 5** Spin-resolved band structure of monolayer Ti$_2$C and Ti$_2$N at different strains. The black and red lines represent the majority-spin and minority-spin bands, respectively. The dashed line indicates the Fermi level at zero eV.

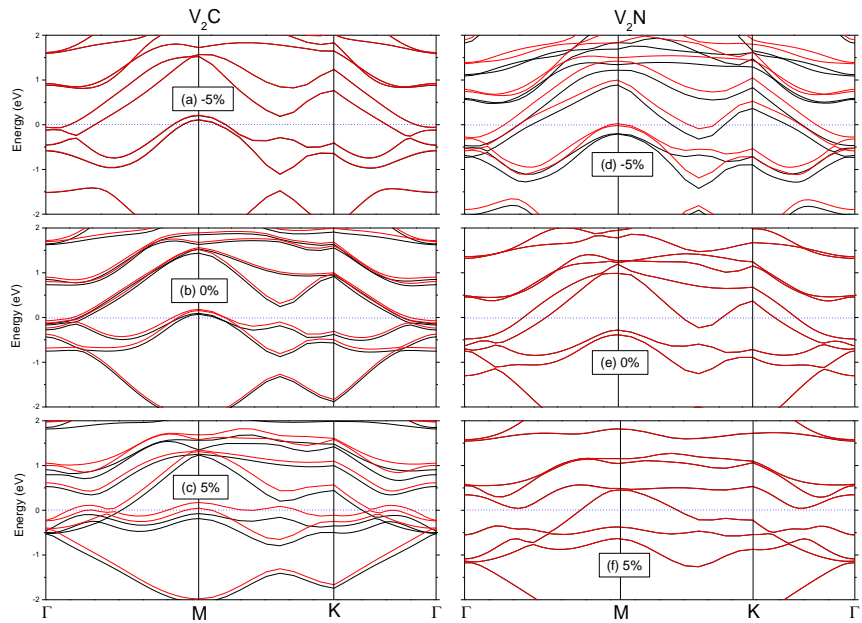

**Fig. 6** Spin-resolved band structure of monolayer V$_2$C and V$_2$N at different strains.

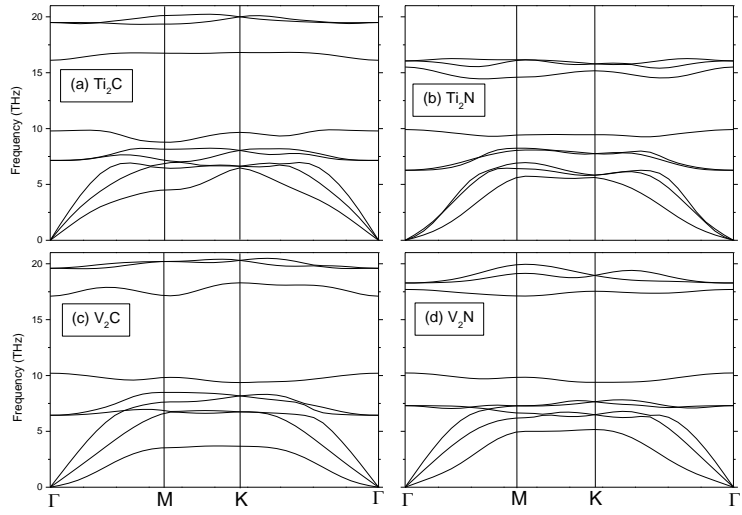

**Fig. 7** Phonon spectrum of monolayer $M_2X$.

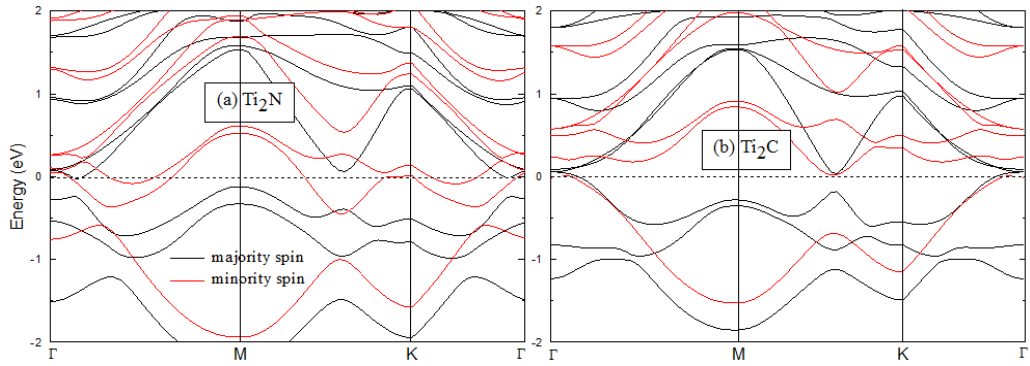

**Fig. 8** Spin-resolved band structure within hybrid (HSE06) functional for monolayer $Ti_2N$ and $Ti_2C$ at the equilibrium lattice constants. The dashed line indicates the Fermi level at zero eV.